\documentstyle[11pt,aaspp4]{article}

\lefthead{}
\righthead{}

\newcommand\hrs{\mbox{$^{\rm h}$}}
\newcommand\mins{\mbox{$^{\rm m}$}}

\newcommand\secsd{\mbox{$^{\rm s} \!\! .$}}
\newcommand\littleprime{\ifmmode{\scriptscriptstle \prime }
	\else{\hbox{$\scriptscriptstyle \prime$ }}\fi}
\newcommand\arcsecpoint{\hbox to 1pt{}\rlap{\arcsec}.\hbox to 2pt{}}
\newcommand\arcminpoint{\hbox to 1pt{}\rlap{\arcmin}.\hbox to 2pt{}}
\newcommand\simlt{\mathrel{\spose{\lower 3pt\hbox{$\mathchar"218$}}
     \raise 2.0pt\hbox{$\mathchar"13C$}}}
\newcommand\simgt{\mathrel{\spose{\lower 3pt\hbox{$\mathchar"218$}}
     \raise 2.0pt\hbox{$\mathchar"13E$}}}

\newcommand\Msun{\hbox{$\thinspace M_{\odot}$}}

\newcommand{\kms}{$\,{\rm km\,s^{\scriptscriptstyle -1}}$}
\newcommand{\gtsim}{\ {\raise-0.5ex\hbox{$\buildrel>\over\sim$}}\ }
\newcommand{\ltsim}{\ {\raise-0.5ex\hbox{$\buildrel<\over\sim$}}\ }

\begin{document}

\title{The Stellar Content of the Halo of NGC~5907 
from Deep HST NICMOS Imaging
\footnote{Based on observations with the NASA/ESA Hubble Space 
Telescope, obtained at the Space Telescope Science Institute, which is 
operated by the Association of Universities for Research in Astronomy, Inc., 
under NASA contract NAS 5-26555.}}

\author{Stephen E. Zepf}
\affil{Department of Astronomy, Yale University, New Haven, CT 06520; \\
zepf@astro.yale.edu}

\author{Michael C. Liu}
\affil{Department of Astronomy, University of California, Berkeley, CA 94720;
\\ mliu@astro.berkeley.edu}

\author{Francine R. Marleau}
\affil{Institute of Astronomy, University of Cambridge, 
Cambridge CB3 0HA, U.K.; marleau@ast.cam.ac.uk}

\author{Penny D. Sackett}
\affil{Kapteyn Institute, University of Groningen,
9700 AV Groningen, Netherlands; psackett@astro.rug.nl}

\author{James R. Graham}
\affil{Department of Astronomy, University of California, Berkeley, CA 94720;
\\ jrg@astro.berkeley.edu}

\begin{abstract}

	We present H-band images obtained with NICMOS of
a field $75''$ (5~kpc) above the plane of the disk of the edge-on 
spiral galaxy NGC~5907. Ground-based observations have shown 
that NGC~5907 has a luminous halo with a shallow radial profile 
between 4 and 8~kpc that roughly traces the dark matter distribution 
of the galaxy deduced from its rotation curve.
Our NICMOS observations were designed 
to resolve bright giants in the halo of NGC~5907 to constrain 
its stellar composition with the goal of understanding 
its nature and origin. More than 100 stars are expected in 
the NICMOS images if the dwarf-to-giant ratio in the halo of 
NGC~5907 is consistent with that expected from standard stellar 
initial mass functions, and if ground-based estimates of the 
distance to NGC~5907 and the integrated colors of its halo are correct. 
Instead we observe only one candidate giant star.
This apparent discrepancy can be resolved by assuming either a 
significantly larger distance than suggested by several studies, 
or a halo metallicity much lower than suggested by ground-based 
colors and as low as metal-poor Galactic globular clusters.
If previous distance and halo color estimates for NGC~5907 are 
correct, our NICMOS results suggest that its extended light
is composed of stars that formed with an initial mass function 
different than that observed locally, leading to a much higher 
ratio of dwarfs to giants. We describe how these three possible
explanations for the absence of bright giants in our NICMOS
images of the halo of NGC~5907 might be constrained by future 
observations.

\end{abstract}


\section{Introduction}

	The presence of dark matter halos around galaxies is
well-established through a number of observations. These include
the flat rotation curves of spiral galaxies, 
the velocities of globular clusters and satellites around their
host galaxies, the properties of hot X-ray gas around ellipticals, 
and mass-to-light ratio measurements from gravitational
lensing (e.g.\ reviews by Sackett 1996, Ashman 1992, Trimble 1987).
However, despite this wealth of data demonstrating the
existence of massive dark halos, their composition remains unknown.  

	The total mass density inferred for galaxy halos is 
roughly $\Omega_{galaxies} \simeq 0.2$ (e.g.\ Bahcall, Lubin, \& Dorman 1995). 
This number comes from the combination of the observed luminosity 
density in the local universe (e.g.\ Loveday et al.\ 1992, 
Efstathiou et al.\ 1988) with the assumption that typical galaxy
halos extend to $\sim 200$~kpc, as suggested by observations
of satellites of spiral galaxies (Zaritsky et al.\ 1993),
and have $(M/L)_{B} \simeq 100 h$ within this radius around 
spirals and a factor of a few higher around ellipticals 
(e.g.\ Mushotzky et al.\ 1994). 
The best constraint on the corresponding value of the
mass density in baryons comes from measurements of deuterium
in quasar absorption lines, from which
$\Omega_{B} h^{2} \simeq 0.02 $ is derived 
(Burles \& Tytler 1998a,b). For current estimates of 
$H_0 = 70~ {\rm km}{\rm s}^{-1}{\rm Mpc}^{-1}$ (Mould et al.\ 1999),
this baryonic density is smaller than the total mass density inferred 
for galaxy halos, indicating that some or most of the dark matter 
around galaxies is in the form of exotic particles. Nevertheless, 
$\Omega_{B}$ is not negligible compared to $\Omega_{galaxies}$,
leaving room for a substantial contribution 
from baryons to the massive dark halos around galaxies. 

        Photometry of edge-on spiral galaxies in search of massive halos 
has been done since the 1970s, originally with TV cameras and photographic
plates (e.g.\ Davis 1975). Faint, extended halo light is easiest to 
detect in thin, edge-on spirals because the regions above and below 
the galactic plane are not contaminated through projection effects 
by the brighter stellar disk.  Skrutskie et al. (1985)
set upper limits on faint $V$-band and, using large apertures, $K$-band 
emission in the late type edge-on spirals NGC~2683, NGC~4244 and NGC~5907, 
concluding that no more than $\simeq 1/3$ of an isothermal dark halo 
could be composed of a luminous baryonic component.  
Refined halo mass models for NGC~5907 based on galaxy parameters derived in  
$H$ and $K$ from pixelated near-infrared detectors and the 
Skrutskie et al. (1985) upper limits for halo light increased 
the baryonic upper limit to $\simeq 2/3$ (Barnaby \& Thronson 1994).

	NGC~5907 is a particularly interesting case since it
exhibits a flat rotation curve well beyond its Holmberg radius 
(Sancisi \& van Albada 1987, Sofue 1994) which allows a good estimate 
of its dynamical mass to be made.  The thin stellar disk is slightly 
warped perpendicular to the line-of-sight.  
In a pioneering CCD photometry study of very low surface brightness 
features around edge-on spiral galaxies by Morrison, Boroson, \& Harding 
(1994), faint extended $R$-band emission was discovered around NGC~5907.  
The extended light is unlike any known thick disk, both in 
terms of its shallow radial profile and its low surface brightness 
(Sackett et al. 1994, Morrison 1999).  However, the faint NGC~5907 
light is well fit with 
a halo-like profile that is moderately flattened toward the plane of the 
galaxy ($c/a \approx 0.5$) and a radial volume density profile 
$\rho \propto r^{-2.3}$, similar to that
inferred for the massive halo from the rotation curve (Sackett et al. 1994). 
A similar flattening has also been tentatively observed in the
globular cluster system of NGC~5907 in a recent study with
HST by Kissler-Patig et al.\ (1999).

	The $R$-band radial profile of the NGC~5907 halo is considerably 
shallower than that of known stellar populations in the halos of the 
Milky Way and M31, which are much steeper than that of the dark halo 
mass inferred from their rotation curves and other dynamical measures. 
Specifically, in the Galaxy the RR Lyrae distribution falls 
off as roughly $r^{-3.5}$ (Saha 1985) and the globular cluster 
system as $r^{-3}$ (Zinn 1985) or $r^{-3.5}$ (Harris \& Racine 1979). 
In M31, the globular clusters are distributed like $r^{-3}$ 
(Racine 1991), while red giant branch stars appear to have a profile 
falling at least as steeply as $r^{-3.8}$ (Reitzel, Guhathakurta, \& 
Gould 1998). In contrast, various kinematical and lensing constraints 
suggest that the total mass distribution of our Galaxy and other spirals 
has a much shallower radial profile, perhaps $r^{-1}$ within a few kpc 
and tapering to roughly $r^{-2}$ out to at least 50~kpc 
and perhaps much beyond (Sackett 1996, 1999 and references therein; 
Zaritsky 1999 and references therein).

	The existence of this unusual stellar component with
a shallow radial profile in NGC~5907 has been confirmed by a number 
of independent observations by other groups, both in the near-infrared 
(Rudy et al.\ 1997, James \& Casali 1998) and at other 
optical wavelengths (Lequeux et al.\ 1996, 1998, Zheng et al.\ 1999).  
Although different proposals have been made for the origin of the 
extended light in these studies, all radial profiles agree at 
surface brightness levels of $R \le 27$ mag/arcsec$^2$, corresponding 
to a height above the plane of about $120''$ (8 kpc for a distance 
of 14 Mpc). 
Lequeux et al.\ report optical colors for this shallow, 
extended population of $(B-V) \simeq 1.0$ and $(V-I) \simeq 1.4$, 
as red as typical elliptical galaxies, while red $J - K$ colors 
of $1-1.5$ have reported in the infrared studies 
(Rudy et al.\ 1997, James \& Casali 1998). 
At longer wavelengths of 3.5 - 5$\mu$m, Yost et al.\ (1999) have 
recently reported a non-detection around NGC~5907 at $180''$ to $540''$ 
(12 $-$ 37 kpc). Combining this result
with model atmospheres, they conclude that hydrogen burning stars 
contribute no more than 15\% of the dark mass within this region. 
Although this non-detection at 3.5 - 5$\mu$m does not coincide spatially
with the detections at $\le 2.2\mu$m, the negative result at longer
wavelengths suggests that whatever produces the faint optical light 
around NGC~5907 does not emit strongly at these wavelengths and probably 
does not have enough mass to account for the dark matter halo inferred 
from the rotation curve.

	The origin of the stellar halo of NGC~5907 raises many puzzles. 
Its shallow radial profile differs from known disk populations 
and red colors are different from known halo populations.  
The relatively red color implies either an initial mass function 
(IMF) favoring extremely low mass dwarfs ($ < 0.2 M_\odot$), 
or a normal, metal-rich, old stellar population. 
Stellar halos are expected to be metal-poor, as observed 
in the Milky Way, because halo formation is believed to occur
early in the life of a galaxy before the majority of metals
have been generated in nuclear processes in stellar interiors.
But an IMF as strongly biased towards low mass dwarfs as is required 
to explain the halo colors of NGC~5907 has not been 
observed in globular clusters or in the local halo population of the 
Milky Way (e.g.\ King et al.\ 1998 and references therein).
On the other hand, if the red colors are generated by an old, 
metal-rich population, one must explain how such a population 
came to reside in the halo where metal enrichment is thought to be low. 

	One possibility proposed by Lequeux et al. (1998) 
is that such a halo could result from accretion of a
metal-rich population from a low-mass elliptical after a tidal encounter
with NGC~5907.  If such an encounter could disrupt the elliptical 
without damaging the thin disk of NGC~5907, the debris may naturally 
settle into a halo configuration similar to what is observed.
Alternatively, Fuchs (1995) suggests the extended halo population
originates from the dynamical response of the stellar spheroid to
the dark matter halo in which it is embedded.
Zheng and collaborators suggest that the faint light seen 
in their deep observations of NGC~5907 --- and by inference the 
observations of others as well --- may be a combination of effects 
including confusion from a very narrow arc apparently associated with the 
galaxy, an unseen but postulated face-on warp, and stellar foreground 
confusion.  If these suggestions, all of which involve normal 
stellar populations, are correct, the surface brightness 
observed from the ground should be associated with a substantial number 
of giants that can be resolved from space.

	In this paper, we present and discuss the implications of 
NICMOS observations of the halo of NGC~5907 
designed to detect the many individual bright giants expected to 
contribute significantly to the halo light of NGC~5907 if it 
is composed of a stellar population with a
initial mass function like that observed locally and has a distance
and colors consistent with existing observations.
In Section 2, we describe the observations, data reduction and analysis
including the detection limits, and show the resulting mosaic of 
our images of the stellar halo of NGC~5907. The comparison of these 
observations to models of the stellar population of this component
is presented in Section 3. This section includes a detailed
discussion of the effects of distance, metallicity, and initial
mass functions on these models and thus the interpretation of our
data. The implications of the absence of stellar
sources in our H-band images of the observed stellar halo of
NGC~5907 are discussed in Section 4.

\section{Observations and Data Reduction}

	We used the near-infrared camera (NICMOS; MacKenty et al.\ 1997) 
on the Hubble Space Telescope (HST) to observe a field in the halo 
of NGC~5907 during two separate observing windows on 1998 May 23 and 
1998 July 10-11. A total 35,326~s of integration time over 12 orbits
was obtained. One dark exposure of zero length was taken 
at the beginning of each orbit in the ACCUM mode
to reduce persistence effects from cosmic rays, and the remainder 
were taken through the F160W filter in the MULTIACCUM mode with the 
SPARS64 sequence.  
Between the two separate observing windows, the position angle of 
the telescope was free to rotate and rolled clockwise by $43.7\deg$.   
Observations were obtained with the NIC2 camera, 
which has a field of view 19\arcsecpoint2 $\times$ 19\arcsecpoint2 wide 
with 256 pixels on a side and 0\arcsecpoint075 per pixel. 
The telescope was pointed so that the geometrical center of the 
NIC2 camera was $\alpha$ = 15\hrs 16\mins 01\secsd54, 
$\delta$ = 56$\deg$ 20\arcmin 31\arcsecpoint7 (J2000). 
This location is 75\arcsec
away from the center of the
galaxy, perpendicular to the plane of the galaxy (Figure 1),
corresponding to a distance above the plane of the
galaxy of 5.1~kpc based on a distance to NGC~5907 of
14~Mpc (see Section 3.3 for a discussion of the distance to the galaxy).
This position was chosen to be outside both the region of disk confusion 
and the thin, long arc found by Shang et al. (1998), but well 
inside the region where extended light has been reported in BVRIJK.

	The images were processed through CALNICA, 
the standard NICMOS pipeline, which performs bias subtraction, 
dark-count correction, and flat-fielding.  
Our own subsequent data reduction consisted of masking bad pixels, 
correcting for constant offsets between the four quadrants of the 
NIC2 camera, and subtracting a scaled master ``sky frame'' constructed
from the average of all of the individual images. The images were
then registered and combined.
Due to the different position angles between visit 1
and visit 2+3, the images of visit 1 were combined separately from visit 2+3.
The images of visit 2+3 were registered with respect to each other using 
the brightest object in the field and co-added to improve the signal-to-noise. 
The first image of visit 3 was not included in the combined image because 
of bad trailing due to the loss of one of our guide stars during that 
exposure. The visit 1 mosaic, generated using the world coordinate system 
in the image, was then combined with visit 2+3, after rotating the
latter by an angle of $43.7\deg$ and registering on the brightest
object in the field.  
The rotation was done after mosaicing because rotation spreads 
bad pixels and cosmic ray hits in a way that it is difficult to 
remove them accurately when combining individually rotated images.
Each resulting mosaic covers a field of view 23\arcsecpoint75 $\times$ 
20\arcsecpoint25 wide.  The final combined F160W image of our field
is shown in Figure~2.  The signal-to-noise ratio is higher at the
center of the image, where the total integration time is longer,
and lower at the edges of the frames.  

\subsection{Photometry}

	Photometry was performed on the final combined image using
the version of the automated star-detection algorithm DAOPHOT 
(Stetson 1992) as implemented in the IRAF package.  
In order to allow a uniform threshold for object detection 
to be applied across the whole image, each pixel in the
image was normalized by the square-root of the exposure time 
at that location. The object detection was performed with
the subroutine DAOFIND and a detection threshold of 5 $\sigma$ 
above the local background level. Objects were also required 
to have DAOPHOT parameters $ -0.5 < $ SHARP $< 0.5 $ and CHI2 $< 5$. 
These cuts are aimed at eliminating features that have a significantly
smaller or larger extent than the PSF (e.g.\ cosmic rays or single pixel 
defects and unresolved blends or large galaxies, respectively). 
Only one object meets these criteria and is a potential star.
This object is also the only object with a FWHM less than
twice the FWHM of the PSF.

	The single stellar object detected is marked on Figure~2
with a circle. The total magnitude of that star is $m_{F160W} = 23.6$.
This is determined by obtaining photometry in an aperture
of 1.7 pixels in diameter and using an aperture correction
derived either from TINYTIM models or NIC2 observations of 
brighter stars in other fields (these give the same results).
The magnitudes in the F160W filter are very similar to typical 
ground-based H-band magnitudes, with an uncertainty in the calibration 
of NICMOS photometry for an object with unknown color of about $10\%$ 
(Colina \& Rieke 1997). 
Due to the high Galactic latitude of NGC~5907 ($b = 51\deg$), 
models of the stellar distribution in the Galaxy yield a negligible 
probability of a foreground star in our field (e.g.\ Cohen et al.\ 1993).
However, there are few constraints on Galactic sub-stellar 
(e.g.\ brown dwarf) populations at these very faint infrared magnitudes.
Based on deep NICMOS pointings in blank areas of the sky
(e.g. Yan et al.\ 1998), we expect roughly ten faint field galaxies 
in the images, consistent with the number of resolved sources we detect.

\subsection{Artificial Star Tests}

	In order to determine the magnitude limit of our photometry,
we performed a series of artificial star tests on the images.
Specifically, we added artificial stars of known magnitude
into the final NIC2 image used for the analysis above, and then
performed the same object detection and photometry on these images
with the artificial stars that was used on the original image. 
Because of the lack of stellar objects in our field, we used 
a PSF determined from other NIC2 data (Marleau et al.\ 1999) 
to create the artificial stars. For each 0.1 magnitude bin,
105 artificial stars were added to the original image with a 
random spatial distribution. The object identification and output 
magnitudes for the artificial stars that were returned by the same 
procedures used on the real data were then compared to the input list.
This was repeated three times for each 0.1 magnitude bin, and the 
results averaged for each bin.
A plot of the recovery fraction 
(completeness) as a function of input magnitude is given in Figure~3. 
As shown in this plot, the average completeness of the photometry over 
the full image is $50\%$ for objects with $m_{F160W} = 24.9$. 

\section{Results}

\subsection{Fiducial Model}

	The primary goal of our work is to constrain the
nature of the halo of NGC~5907 by comparing the observed
star counts in our NICMOS image to the star 
counts expected for various models of the stellar population
of the halo of NGC~5907.
Our procedure is to take a luminosity function 
in F160W for a given stellar population, either from observations 
such as those in the Galactic bulge, or from theoretical models, 
place this population at the distance of NGC~5907, and normalize 
the luminosity function to match the observed surface brightness 
within the region of our NICMOS data. We then
convert this into a prediction of the number of stars
expected in our NICMOS images by multiplying the predicted
distribution in apparent magnitudes with the completeness
function determined above. 

	We begin by considering a fiducial model, based
as closely as possible on the observed properties of NGC~5907
and its extended light distribution. Specifically, our fiducial
model has a surface brightness within the NICMOS region of 
$R = 25.85$ mag/arcsec$^2$ as measured by Morrison et al.\ (1994), 
a distance to NGC~5907 of 14 Mpc based both on
Tully-Fisher and models for deviations from the Hubble flow
in the region around NGC~5907 (see Section~ 3.3), 
and an H-band stellar luminosity function adapted from studies 
of Baade's window (Tiede et al.\ 1995), which has a metallicity 
similar to that suggested by the observed broad-band colors 
of the extended stellar light in NGC~5907.
This fiducial model predicts that more than 100 stars should be 
seen in the combined NICMOS image, in stark contrast to our observation 
of one candidate star. This is graphically demonstrated in
Figure~4, in which we simulate the expected appearance of
the fiducial model in our NICMOS image, accounting for the
expected Poisson shot noise. The significant number of bright 
giants expected for the simplest assumptions about the stellar
population in the halo of NGC~5907 are not present in the 
data. Because of the stark difference between the observations and
the simplest prediction, we consider in turn each of the components 
that go into the prediction. A fundamental component of our calculation 
is the H-band luminosity function, which depends on the initial mass 
function (IMF), metallicity, and age of the stellar population. 
The distance to NGC~5907 and the surface brightness of its halo 
within our NICMOS field also play a role and we consider all 
of these below.

\subsection{Luminosity Function, IMF, and Metallicity}

	For the fiducial case shown in Figure 4, we adopt
an H-band stellar luminosity function constructed from the
K-band luminosity function observed in Baade's Window 
(Tiede et al.\ 1995), adjusted to F160W by a small
color term. The choice of a luminosity function based on
the Galactic Bulge is motivated by the similar metallicities 
of stars in Baade's Window (e.g.\ McWilliam \& Rich 1994)
and the halo of NGC~5907, as inferred from its optical colors
as described in detail below. This approach also has the
benefit of comparing the data to an empirical luminosity
function. As demonstrated in Figure 4, the luminosity function 
based on observations in Baade's Window dramatically fails 
to account for the data. 

	Because of the failure of the fiducial model,
we explored a range of stellar initial mass functions
and metallicities to search for sets of parameters that are 
consistent with the data. We generated theoretical F160W 
luminosity functions using the single burst 12 Gyr-old stellar population
synthesis models of Bruzual \& Charlot (1998) with the semi-empirical
stellar spectral energy distributions of 
Lejeune, Cuisinier, \& Buser (1997). 
The models are defined for [Fe/H] = ${-2.3, -1.7, -0.7, -0.4, 0.0, +0.4}$
with a stellar mass range of $0.1-125$\Msun . We also consider
a range of stellar initial mass functions, parametrized as
a single power law with slope $\alpha$, such that
$dN = M^{-(\alpha+1)} dM$, where a Salpeter IMF is $\alpha = 1.35$. 
We caution that the stellar population models become unreliable
for $\alpha \gtsim 4$ because at these very steep slopes the
stars at the bottom of the main sequence become the principal 
contributors to the integrated light. In this case, the accuracy
of the model predictions rests on the assumed colors of old late-type
M dwarfs and the shape of the IMF near the hydrogen burning limit,
both of which are poorly constrained locally, much less as a function
of metallicity. Nevertheless, the qualitative behavior of the models
for cases of extremely large $\alpha$ should be correct, even if
they are more uncertain quantitatively.


	Figure~5 shows H-band luminosity functions predicted for 
our NICMOS observations of the halo of NGC~5907 as a function of
IMF slope and metallicity, given the ground-based observations 
of the R-band surface brightness of $25.85$ magnitudes/arcsec$^2$
(\S3.4), a distance to NGC~5907 of 14 Mpc (\S3.3),
and the completeness function shown in Figure~3 (\S2.2). 
Figure~5 shows that only extremely steep IMFs or
very low metallicities are consistent with our NICMOS data
in which only one possible star is detected. In the former case,
the absence of giants originates in an extraordinarily
higher dwarf to giant ratio in the stellar halo of NGC~5907,
while in the latter, low-metallicity giants are simply too faint 
to be detected because of the dependence on metallicity of the 
brightness of giants in the H-band.

	This result can be quantified further by considering the
Poisson statistics of either one or zero detected stars.
For metallicities of [Fe/H] $\ge -0.7$, even the largest $\alpha$ 
we consider is discrepant with the observation of one star
at $> 99.99\%$ confidence level. Because of the uncertainties in 
the stellar populations models in these extraordinarily dwarf 
dominated  IMFs discussed above, we simply adopt $\alpha > 4$ 
as the result for [Fe/H] $\ge -0.7$. Alternatively, if [Fe/H] is
sufficiently low, then the brightest giants become too
faint to detect. If one adopts the usual Salpeter IMF
with $\alpha = 1.35$, this requires [Fe/H] $\ltsim -1.7$.
As can be seen in Figure~5, the transition from the presence
of giants at [Fe/H] $\ge -0.7$ to the absence of giants at
[Fe/H] $\ltsim -1.7$ is fairly insensitive to modest changes 
in $\alpha$. We also note that if the light in NGC~5907 is 
assumed to come from a solely low metallicity population, 
the one stellar object is unlikely to be a star in the halo 
of NGC~5907, since the low metallicity 
models predict that any stars that are detected are found at the 
faintest limit of the data, while the stellar object is more 
than one magnitude brighter than our $50\%$ completeness limit.


	Tighter constraints on the stellar population of
the extended light around NGC~5907 can
be placed by requiring the integrated colors of the
halo population to be consistent with the ground-based
colors of the halo light reported by Lequeux et al. (1998).  
Figure~6 compares the observed (B$-$V) and (V$-$I) colors 
to the BC98 model populations over the same range of $\alpha$ 
and [Fe/H] shown in Figure~5. With the NICMOS data alone,
it is possible to account for the absence of giants through
very low metallicity since then the giants would be too faint
to detect in our images. Figure~6 shows that the published
optical colors argue against the possibility of a 
very low metallicity with a normal IMF.
We also note that the near-infrared (J$-$K) colors
of Rudy et al.\ (1997) and James \& Casali (1999) are even
redder than the models that can account the optical colors,
which are already red. While it is unclear whether the problem
is with these challenging observations, or with difficulties
in the models for either the reddest metal-rich
giants (important for populations with normal IMFs) or
the latest type M dwarfs (important for the steep IMFs), 
it is clear that the observational evidence to date 
suggests that the colors of the NGC~5907 halo are red. 
Therefore, the joint NICMOS star count and color
constraints appear to require a dramatically higher dwarf-to-giant
ratio than given by a typical IMF.

	More quantitatively, in order to be consistent within the $99\%$
confidence limit of our observation of only a single star 
in the NICMOS field, the ratio of bright giants to fainter 
stars must be more 100 times lower than that expected for a stellar 
population with a Salpeter IMF and a metallicity that is consistent
with the optical colors at the $2\sigma$ level.
This constraint translates to $\alpha > 3$ 
for a simple power-law parametrization of the IMF.
Such steep IMFs cause low metallicity models which are consistent 
with the absence of bright giants ([Fe/H] $\ltsim -1.7$)
to become red enough to begin to match the optical colors. 
Because this result indicates a much higher
ratio of dwarfs to giants than observed in Galactic globular
clusters or expected from the stellar initial mass function
in star forming regions of the Galaxy, we examine the robustness 
of each of the steps taken in obtaining this result.

\subsection{Distance}

	The detection of bright giants in NGC~5907 clearly
depends on the distance to the galaxy. Using both the H and R-band 
Tully-Fisher relations, as well as the observed radial velocity of NGC~5907 
combined with a model of the expected peculiar velocity at its 
location, we find that NGC~5907 is at a distance of 14~Mpc, with 
an uncertainty of $20\%$. Half of this uncertainty is due to
the intrinsic scatter of the techniques we use to determine
the distance to NGC~5907. The other half is due to the current
uncertainty in the extragalactic distance scale, as manifested 
in the uncertainty in the distance to the Virgo cluster or similarly 
in the Hubble constant. This comes primarily from the uncertainty 
in the absolute calibration of the Cepheid Period-Luminosity relation.
A distance to NGC~5907 of 14~Mpc is somewhat greater than that 
typically adopted in earlier work (e.g.\ Morrison et al.\ 1994),
and our own preliminary presentation of this work at conferences
(Liu et al.\ 1998). 

\subsubsection {H-band Tully-Fisher}

	The H-band Tully-Fisher relation is one of the most
reliable methods for determining distances to edge-on spiral 
galaxies like NGC~5907. The Tully-Fisher relation
has been heavily studied and tested, resulting in well-established
relationships between spiral galaxy luminosity and line-width
(e.g.\ Jacoby et al.\ 1992). Moreover, the H-band offers a significant 
advantage in its reduced sensitivity to internal extinction, which 
is the dominant source of uncertainty for optical determinations of the 
magnitudes of edge-on galaxies. Even in H, the internal 
extinction is probably not zero for galaxies as inclined as NGC~5907 
(e.g. Moriondo, Giovanelli, \& Haynes 1998, Tully et al.\ 1998). 

	One way to determine the H-band Tully-Fisher distance
to NGC~5907 is to adopt the H-band Tully-Fisher relation given by Hubble
Space Telescope Key Project on the Hubble Constant, which is based 
on Cepheid distances to 21 spiral galaxies (Sakai et al.\ 1999). 
Specifically, the Key Project team found an H-band Tully-Fisher 
relation of $H^{c}_{-0.5} = -11.03({\rm log} W^{c}_{20} - 2.5) - 21.74$,
where $H^c_{-0.5}$ is the H-band magnitude within an aperture
that is a fixed fraction of the B-band diameter, 
as defined by Aaronson, Huchra, \& Mould (1979),
and $W^c_{20}$ is the velocity width of HI at $20\%$ 
of the maximum HI flux, corrected for inclination.
The observed dispersion in the relation is 0.36 magnitudes for 
the Key Project sample, similar to that found for samples of 
cluster spirals (e.g.\ Peletier \& Willner 1991, 1993; 
hereafter PW91 and PW93).
The slope of this H-band Tully-Fisher relation is similar to,
but slightly larger than that found for spirals in Virgo and Ursa Major 
found by PW91 and PW93, 
and by the earlier work of Aaronson, Huchra, \& Mould (1979). 
Thus, the Sakai et al.\ relation will give a slightly larger
distance for NGC~5907, which has a larger velocity width than
most galaxies in the sample, although the difference is well
within the 0.36 magnitudes internal scatter. 
	
	For NGC~5907 itself, we adopt the values of $H_{-0.5} = 7.58$ 
and log $W_{20}^{c} = 2.690$ given by Aaronson et al.\ (1982).
For NGC~5907, the value of log $W_{20}^{c} = 2.690$ is well-determined
for this very edge-on system (e.g.\ Sch\"oniger \& Sofue 1994). 
The H magnitude of NGC~5907 may be more uncertain, as the galaxy 
is edge-on and internal extinction is likely to play a role even 
in the H-band. Two recent studies have attempted to determine the 
effect of extinction for spiral galaxies in the near-infrared. 
Based on a large study 154 spiral galaxies, 
Moriondo, Giovanelli, \& Haynes (1998) find 
an extinction in H of approximately 0.15 magnitudes for a galaxy 
with the inclination of NGC~5907. A smaller study by Tully et al.\ (1998) 
in K, finds 0.25 magnitudes of extinction in the K-band. 
We therefore adopt 0.2 magnitudes
as the best estimate of the H-band extinction in NGC~5907,
and note that the $\sim 0.4$ magnitudes scatter in the Tully-Fisher
is sufficient to account for the uncertainties in this
correction. With log $W_{20}^{c} = 2.690$ 
and $H_{-0.5}^{c} = 7.38$, we find that the distance to
NGC~5907 is 31.21, or 17.5 $\pm 2.7$ Mpc, based on the Key Project relation.

	An alternative is to use the slope found for spirals
in Ursa Major and Virgo while still setting the zero point
to give the same 16 Mpc distance to the Virgo cluster
determined by HST Cepheid observations (Macri et al.\ 1999). 
Peletier \& Willner studied roughly 25 spirals in each of these 
clusters and found a slope of 
$H^{c}_{-0.5} = -10.2({\rm log} W^{c}_{20} - 2.5)$.
We adopt the same zero point as above for the Key project sample,
effectively setting the two Tully-Fisher relations equal
for galaxies with log $W^{c}_{20} = 2.5$. This approach gives a 
distance to NGC~5907 of 16.5 Mpc. The difference is due to the 
somewhat shallower slope found for the cluster surveys

	A third approach is to take advantage of more
modern two-dimensional H-band photometry of NGC~5907 
(Barnaby \& Thronson 1992). Although most calibration work 
is done with the Aaronson et al.\ (1982) aperture data, 
PW91 define a Tully-Fisher relationship
based on the H-band light within the isophote at which the
H-band surface brightness is brighter than 19 magnitudes/arcsec$^2$
($H_{19}$). They find a Tully-Fisher relation for this isophotal
magnitude of log $(V_{20}) = -0.098 (H_{19} - 9) + 2.525$,
with the zero point set by $d_{Virgo} =$ 16~Mpc.
From two-dimensional photometry, Barnaby \& Thronson (1992)
derive face-on surface brightnesses and scale lengths for NGC~5907.
Since the $H_{19}$ magnitude of PW91 is defined within an isophote 
on the sky, it is dependent on galaxy inclination, with more light 
being encompassed for edge-on systems. We therefore translate
the Barnaby \& Thronson model to $i = 60\deg$, which is the most
likely inclination in a random sample of galaxies. In this case,
the $H_{19}$ magnitude gives a distance of 12.9~Mpc based on
the calibration above. Lower and upper limits to the distance
derived from this technique can be derived by translating to edge-on
and face-on system, giving a range in distances from 
11.2~Mpc to 16.1~Mpc.

\subsubsection {R-band Tully-Fisher}

	A somewhat independent check on the distance derived
above can be obtained by considering the R-band rather then
the H-band in the Tully-Fisher method. Morrison et al.\ (1994)
estimated the reddening-free total magnitude of NGC~5907 of 
$m_R = 9.1$ by modelling the disk outside of the dust lane, 
and then extrapolating this model into the masked regions.
Taking the R-band Tully-Fisher calibration of Pierce \& Tully
(1992, see also Jacoby 1992) and the rotation velocity
given above, gives a distance of 14.3~Mpc. 
This R-band distance estimate is consistent with the H-band distance 
given above within the uncertainties of about 0.36 magnitudes
for each technique.

\subsubsection {Peculiar Velocity and Observed Density Field}

	The scatter in Tully-Fisher is not negligible, so
it is important to consider other constraints on the distance
to NGC~5907. One approach is to use the observed redshift
of NGC~5907, and the expected peculiar velocity for a
galaxy at its location to estimate its true distance.
This requires a model for the underlying density field,
which has been estimated both from IRAS galaxy surveys
(Davis, Nusser, \& Willick 1996) and optical galaxy surveys
(Baker et al.\ 1998), using the expansion technique of 
Nusser \& Davis (1994). The predicted peculiar velocity for 
NGC~5907 is then --250 \kms\ from the optical survey and --190 \kms\ 
for the IRAS survey. For field galaxies like NGC~5907, the
uncertainty in these peculiar velocities is dominated by
the random dispersion of galaxies around the smooth Hubble Flow,
which is measured to be 120 \kms\ (e.g.\ Baker et al.\ 1999).
Combined with $cz = 667$ \kms\ for NGC~5907 (RC3, de Vaucouleurs
et al.\ 1991) and $H_0 = 70$ \kms\ Mpc$^{-1}$ for consistency with
the Virgo distance above, we find distances of 13.1 and
12.2 Mpc respectively, with an uncertainty of $ \pm 1.7$ Mpc
due to random motions of galaxies. We account for the additional
systematic uncertainty introduced by the Hubble Constant below.

\subsubsection {Final Distance Estimate}

	To determine the distance to NGC~5907, we combine the
Tully-Fisher and flow-field estimates in quadrature. For Tully-Fisher,
we simply average the $H^c_{-0.5}$, $H_{19}$, and R-band estimates,
for which we find 14.9~Mpc. Since the errors in these estimates
are highly correlated, we retain an internal error of 0.36 magnitudes, 
or $\pm 2.3$ Mpc. 
Similarly, for the flow-field estimate, we simply average the
results from the IRAS and optical galaxy predictions, which
gives a distance of $12.7 \pm 1.7$ Mpc. We then combine these
in quadrature, obtaining an answer of $13.5 \pm 1.4$ Mpc.
Both techniques depend directly on the Hubble constant,
for which we have adopted a value of $70$ \kms Mpc$^{-1}$
(e.g. Mould et al.\ 1999).  We take the uncertainty in this
value to be $12\%$ accommodating the latest results on the 
calibration of the Cepheid Period-Luminosity relationship
(e.g.\ Maoz et al.\ 1999).
Combining this uncertainty in the overall calibration of
the distance scale with the uncertainties intrinsic to the
techniques applied to NGC~5907, we find that the distance to
NGC~5907 is $13.5 \pm 2.1$ Mpc. For most of the work in this
paper, we round this up to 14~Mpc.

	The model luminosity functions in Figures~4 and 5 are
based on this distance. In order for the detection of at most
one star in our NICMOS image to be consistent with a Salpeter
IMF and within $2\sigma$ of published optical colors, NGC~5907
would have to be at a distance of more than 24~Mpc, which is
$5~\sigma$ larger than our estimated distance of $14 \pm 2$ Mpc. 
We also note that the distance we derive is larger than previously 
published distances, so accounting for the absence of bright
giants in the extended light around NGC~5907 would require
even larger discrepancies with earlier work.

\subsection{R-band Surface Brightness}

	The predicted star counts are normalized so that
their integrated flux produces the observed R-band surface brightness;
therefore uncertainties in the surface brightness propagate to 
uncertainties in the predicted NICMOS counts. The R-band
surface brightness within our NICMOS pointing is 
$25.85 \pm 0.15$ mag/arcsec$^2$ based on a direct measurement
from the reduced R-band image of Morrison et al.\ (1994), kindly
provided by H.~Morrison and J.~Monkiewicz.
Independent measurements of the surface brightness
in narrower filters at roughly the same wavelengths give good
agreement on the surface brightness of the diffuse stellar
component out to $R < 27$ mag/arcsec$^2$ (Zhang et al.\ 1999).
Thus, the uncertainty in the surface brightness within our
NICMOS field is small compared to other uncertainties and 
to the dramatic difference between the number of giants expected 
and the number observed in the NICMOS image.

\section{Conclusions}

	The fundamental result of our investigation
is that we detect only one unresolved object in our NICMOS
images of the stellar halo of NGC~5907, compared to more than
100 giant stars that are expected to be observable 
within this field, given the observed surface brightness and 
colors in the region of our pointing and the simplest assumptions
about the stellar population of the halo and the distance
to the galaxy. Taken at face value, this result indicates
that the dwarf to giant ratio is about 100 times
greater than that for typical stellar populations.

	Given the dramatic nature of this result, we consider
other options. A very low metallicity ([Fe/H] $\ltsim -1.7$)
results in giants too faint to be seen by our observations.
However, a stellar population with such a low metallicity and
a Salpeter IMF has colors for the integrated light that are 
more than 3$\sigma$ bluer than the $(B-V)$ and
$(V-I)$ observations of Lequeux et al.\ (1998). Alternatively,
if NGC~5907 is more distant than typically believed, then 
the giants may be too faint to be observed. However,
for a stellar population with a Salpeter IMF and a metallicity 
that gives colors within $2\sigma$ of the observed optical colors, 
the distance must be more than 24 Mpc to be even marginally
consistent with our detection of only one stellar object.
This is $5\sigma$ greater than our estimate of $14 \pm 2$ Mpc
for the distance to NGC~5907, and even more discrepant with
earlier work, which adopted closer distances (e.g.\ Morrison et al.\ 1994,
Yost et al.\ 1999). Thus, only the ``last resort'' option of a stellar 
population with a very high dwarf-to-giant ratio appears to be 
able to account for the absence of resolved stars in our NICMOS image 
in the stellar halo of NGC~5907 without conflict with other observational 
data. Specifically, for a simple power-law parametrization of the IMF,
$\alpha > 3$ is required to be within $2\sigma$ of the published (B$-$V)
and (V$-$I) colors and within the 99\% confidence limit of the detection
of at most one star.

	If confirmed, this result has important
implications for the nature of the diffuse light observed at 
$z$-heights of 4--8 kpc in NGC~5907. Firstly, our observations do not 
support a number of proposed origins for this stellar component.
Specifically, if this diffuse light originated from an accreted 
elliptical galaxy (Lequeux et al.\ 1998), was dynamically heated 
from a thinner stellar disk, or is a tidal warp or ring
(Shang et al.\ 1998), many giants would have been observed in our 
NICMOS images of NGC~5907, since none of these populations is expected 
to have an extremely steep IMF or a very low metallicity. Although 
a larger distance to NGC~5907 would allow any model of the extended 
stellar light in NGC~5907 to be reconciled with the data, no extant
model provides an {\it a priori} explanation why several different 
techniques should underestimate the distance to NGC~5907. 
Very speculatively, if a high dwarf-to-giant 
ratio is the source of the absence of giants in our NICMOS
images, and the IMF is assumed to be a power-law extended 
to masses less than $0.1 \Msun\ $, then the R-band mass-to-light 
ratio is high enough that the observed halo can constitute 
the mass that produces the galaxy's rotation curve. However, 
such a model is inconsistent with the combination of a non-detection 
of the halo of NGC~5907 at 3.5 -- 5$\mu$ and current models of
stellar atmospheres of stars at the hydrogen burning limit 
(Yost et al.\ 1999).

	Because a dramatically dwarf-rich stellar population 
has not been observed elsewhere, we consider several additional 
ways to constrain the nature of the observed stellar halo of NGC~5907.
Although a stellar population with low metallicity and a standard IMF 
is not consistent with the published optical and near-infrared colors,
measurements of colors at faint surface brightnesses are difficult.
An independent, direct test of the low metallicity hypothesis can be 
made through WFPC2 observations. These should see thousands of
giants if the observed halo light in NGC~5907 comes from
metal-poor stars with a standard IMF and the distance of the
galaxy is consistent with existing estimates. However, if the
WFPC2 observations yield a null result, it will be difficult to 
obtain further independent tests of the distance to NGC~5907.
One possibility is to obtain NICMOS observations closer to the 
center of the galaxy, which should reveal some bright stars
if the distance to the galaxy is not much greater than expected.
Initial HST observations of its globular cluster system did not
reveal a sufficient number of clusters to reliably use the
peak of the globular cluster luminosity function as a
distance indicator (Kissler-Patig et al.\ 1999). 
Planetary nebulae might be observable, although observations
in the halo could be suspect if the IMF is as dramatically
different as suggested here. 

\acknowledgments 

	We thank St\'ephane Charlot for providing us with the
isochrone synthesis data for galaxy evolution. We thank Heather
Morrison and Joe Silk for valuable discussions, and the referee
for helpful suggestions.
We also thank Heather Morrison and Jackie Monkiewicz
for providing direct measurements of the optical surface brightness 
in their images within our NICMOS field. 
We acknowledge excellent assistance from Doug van Orsow at STScI 
in getting our program executed and financial support from the following 
HST NASA grants GO-07277 (SEZ and FRM) and AR-07523 (FRM). 
FRM also acknowledges an IoA postdoctoral research fellowship.

\clearpage

\begin{figure}
\plotone{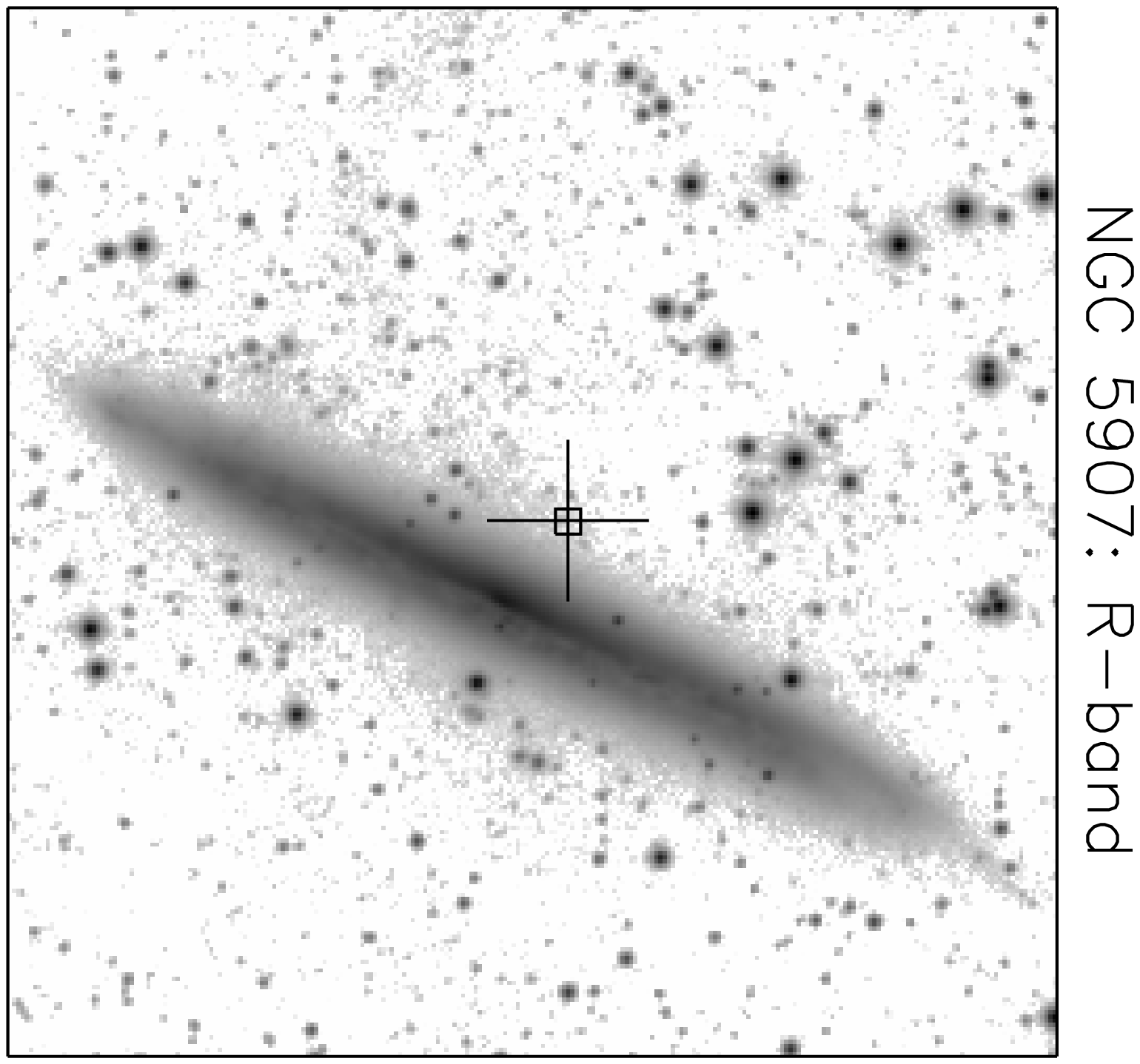}
\vskip -1.0in
\caption{A deep R-band image of NGC~5907 from 
Morrison et al.\ (1994), with our NICMOS field
overplotted as a square. The image is 12\arcminpoint8
on a side, and the NICMOS pointing is $75''$ above
the plane of the disk. North is up and east is to
the left.}
\end{figure}

\begin{figure}
\plotfiddle{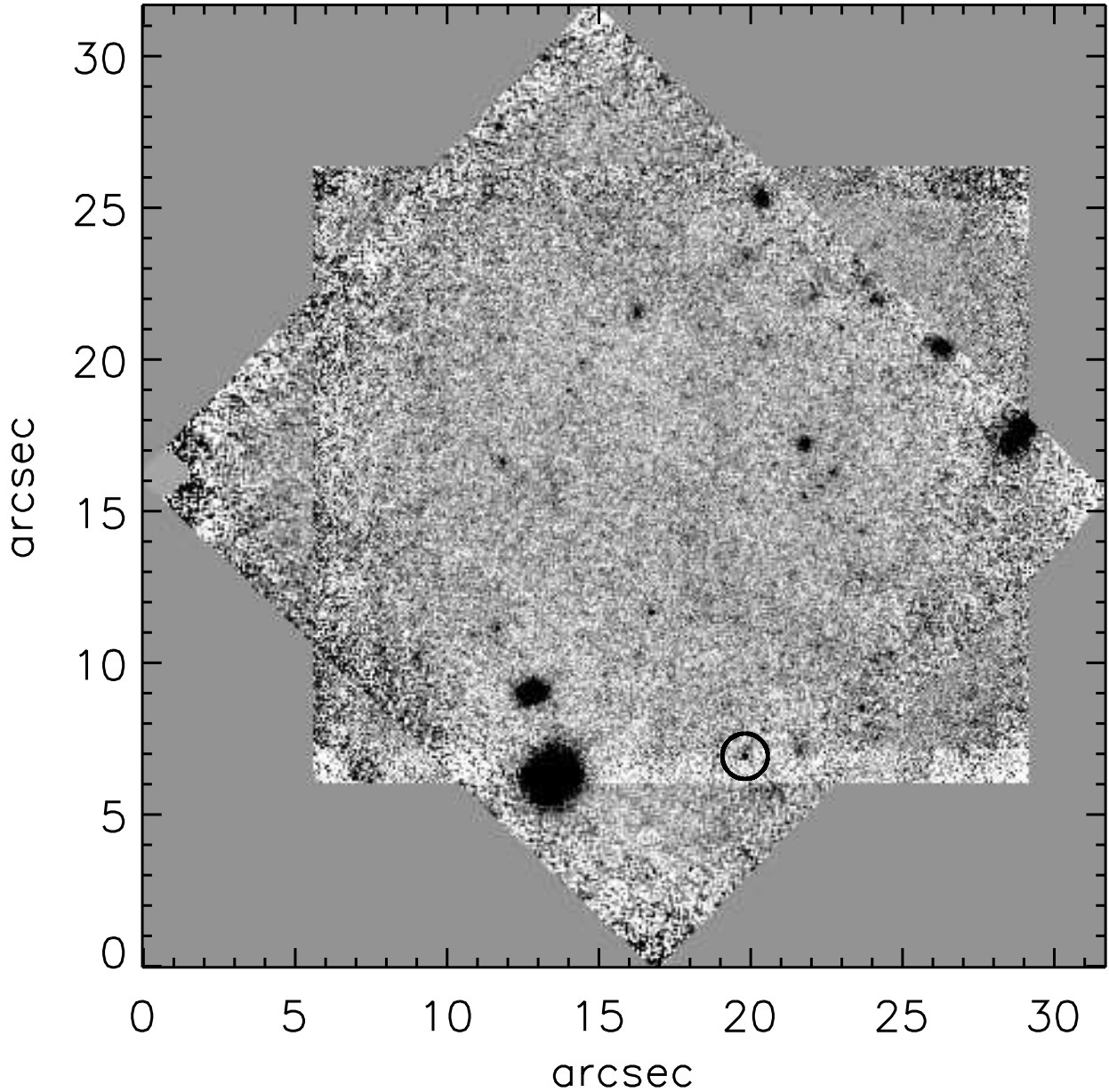}{7.5in}{90}{100}{100}{460}{0}
\vskip -0.4in
\caption {The NICMOS/F160W mosaic of our field in the halo of NGC~5907. 
The data show a single unresolved object (circled on the image)
with $m_{F160W}= 23.6$ mag. The handful of other objects seen in 
the image are extended and thus identified as background galaxies.}
\end{figure}

\begin{figure}
\plotone{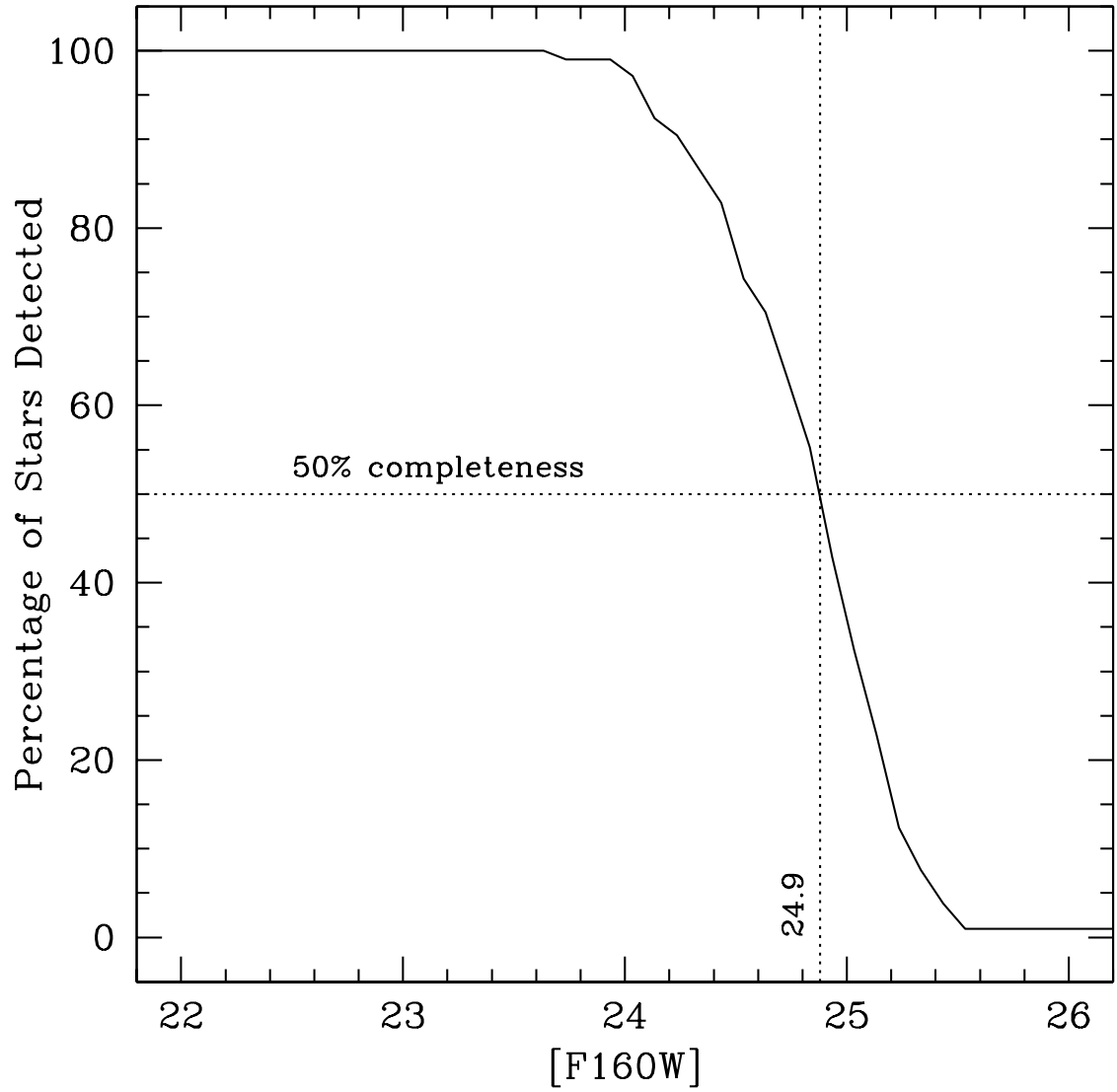}
\vskip -0.1in
\caption{A plot showing the completeness of our photometry
as a function of magnitude.}
\end{figure}

\begin{figure}
\plotone{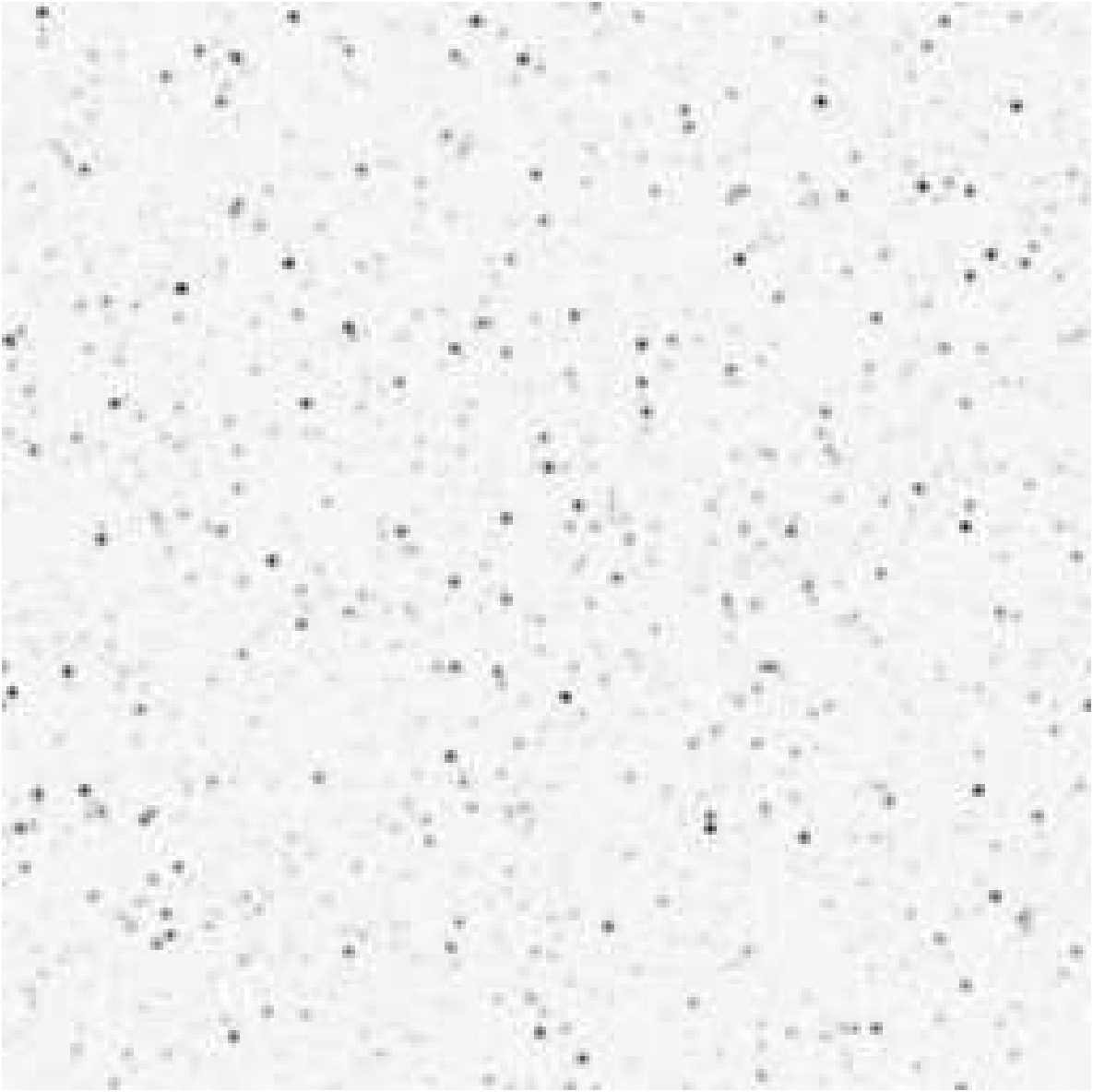}
\vskip -0.1in
\caption{A simulation of the expected appearance of our NICMOS
images of the NGC~5907 halo assuming a stellar population
like that of the Galactic bulge, as suggested by the red optical
and near-infrared colors of the integrated light, and normalized 
to the observed integrated surface brightness of Morrison et al.\ (1994). 
This simulated image 
is 19\arcsecpoint2 on a side.}
\end{figure}

\begin{figure}
\plotfiddle{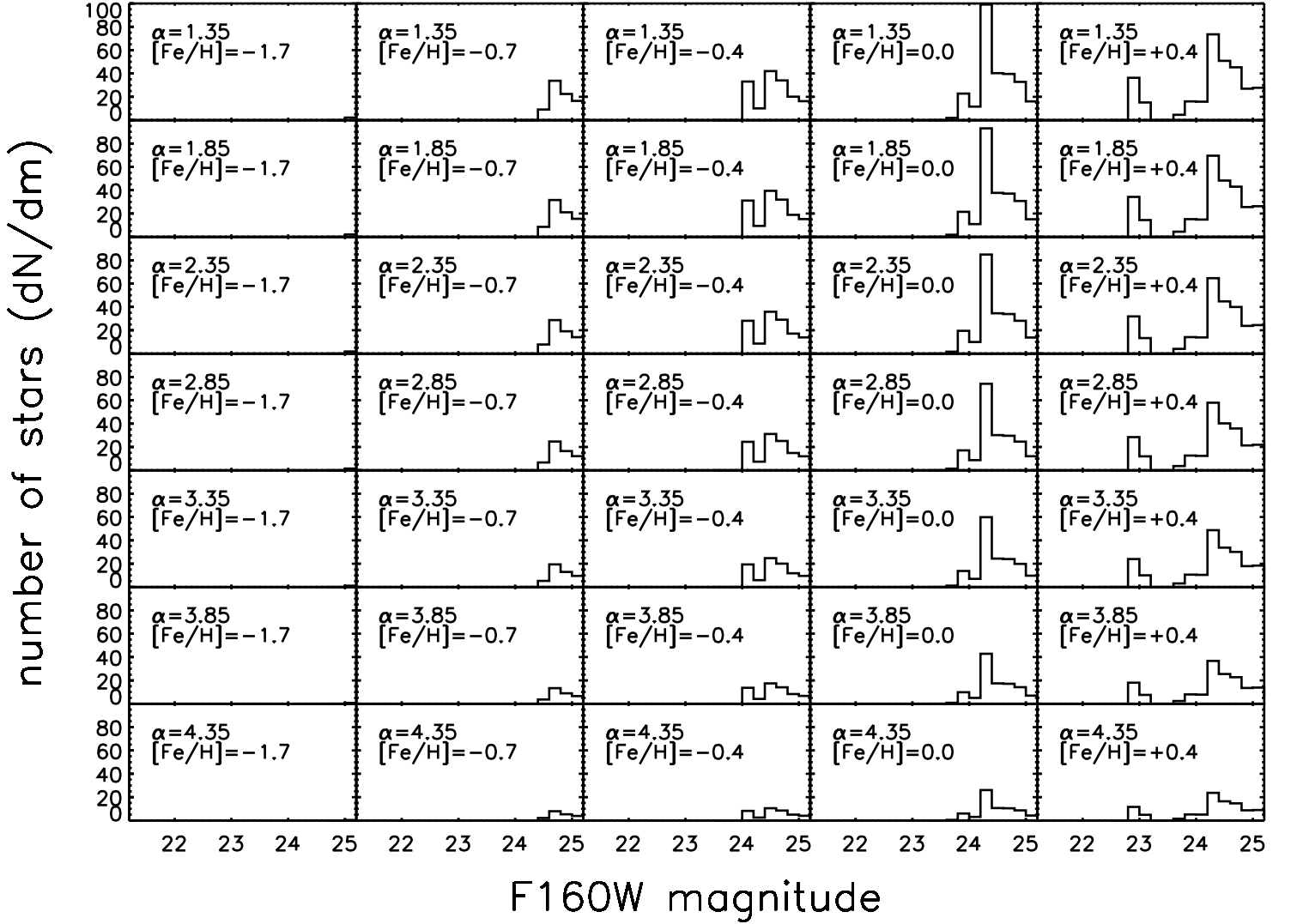}{7.5in}{90}{88}{88}{355}{65}
\vskip -1.25in
\caption{The predicted luminosity functions for our NICMOS
observations. The predicted number of stars as a function of 
magnitude is based on a stellar population model with the given 
initial mass function and metallicity, convolved with the completeness
function given in Figure 3. Most models predict several tens to
hundreds of stars should be observed in our NICMOS field, but
we observe only one potential star, with a magnitude of
$m_{F160W} = 23.6$. This observation allows only models 
of the stellar population within our NICMOS field in the
halo of NGC~5907 with either very low metallicity ([Fe/H $\ltsim -1.7$)
or a very steep IMF ($\alpha > 4.35$).}
\end{figure}

\begin{figure}
\plotfiddle{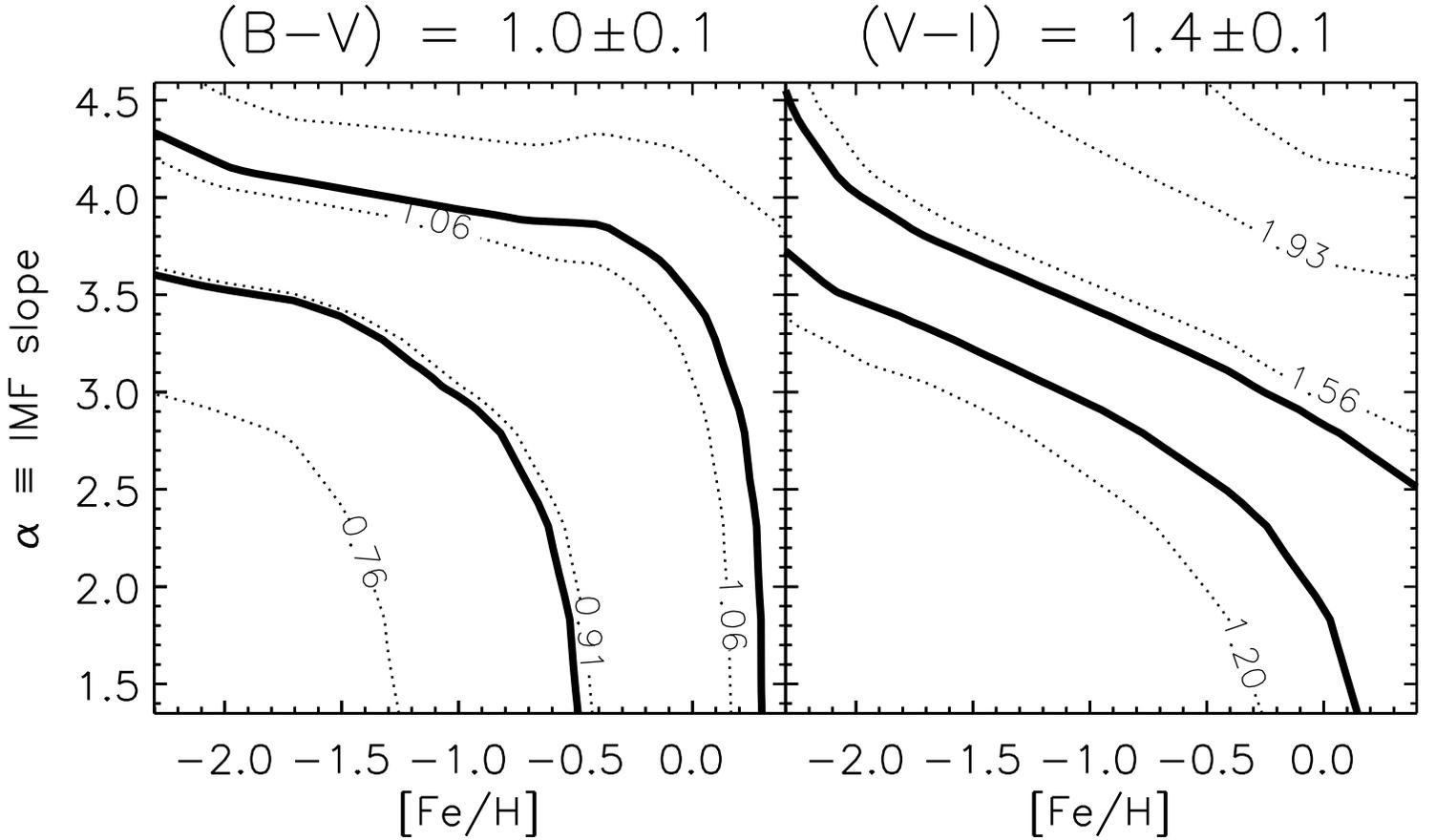}{7.5in}{90}{88}{88}{355}{65}
\vskip -1.5in
\caption{Constraints on the stellar population from the observed
integrated colors of the halo of NGC~5907. The colors and the
errors are from Lequeux et al.\ (1998). The range of IMF slope
$\alpha$ and [Fe/H] are the same as those used in Figure 5.
The light contours show the trends in this diagram for fixed
color. The solid line represents the $1\sigma$ limit of the published
colors. This plot shows that the published colors are strongly 
inconsistent with a low [Fe/H], normal $\alpha$ stellar population 
that might otherwise account for the absence of giants in the NICMOS 
images.}
\end{figure}

\end{document}